\documentclass[12pt]{article}
\usepackage{amsfonts,amssymb,epsfig,amsmath,mathtools,xcolor}
\usepackage{color}

\usepackage{mathtools}
\usepackage{amssymb}
\usepackage{mathrsfs}
\usepackage{graphicx}
\usepackage{subcaption}
\usepackage{hyperref}

\renewcommand{\baselinestretch}{1.2}

\DeclareMathOperator{\Tr}{Tr}
\DeclareMathOperator{\tr}{tr}

\usepackage{graphicx}
\usepackage{epstopdf}
\usepackage{amsmath, amssymb}
\usepackage{comment}
\allowdisplaybreaks

\usepackage{subcaption}
\usepackage{float}
\usepackage{cellspace}
\usepackage{longtable,makecell}
\usepackage{tabularx}
\usepackage{ytableau}
\usepackage[vcentermath]{youngtab}

\usepackage{tablefootnote}
\usepackage{hyphenat}
\ytableausetup{smalltableaux, aligntableaux = center}

\usepackage{multirow}

\newcommand{\be}{\begin{eqnarray}}
\newcommand{\ee}{\end{eqnarray}}
\newcommand{\nn}{\nonumber}
\newcommand{\bn}{\begin{enumerate}}
\newcommand{\en}{\end{enumerate}}

\def\CO{{\cal O}}

\def\L{\Lambda}

\begin{document}

\makeatletter \@addtoreset{equation}{section} \makeatother
\renewcommand{\theequation}{\thesection.\arabic{equation}}
\renewcommand{\thefootnote}{\alph{footnote}}

\begin{titlepage}

\begin{center}
\hfill 
\\

\vspace{2cm}

{\Large\bf Quantum black hole cohomologies}

\vspace{2cm}

\renewcommand{\thefootnote}{\alph{footnote}}

{\large Seok Kim$^1$, Seongmin Kim$^1$, Siyul Lee$^2$ and Jiyoo Park$^1$}

\vspace{0.7cm}

\textit{$^1$Department of Physics and Astronomy \& Center for
Theoretical Physics,\\
Seoul National University, Seoul 08826, Korea.}

\vspace{0.2cm}

\textit{$^2$Instituut voor Theoretische Fysica, KU Leuven, 
Celestijnenlaan 200D, 3001 Leuven, Belgium.}

\vspace{0.7cm}

E-mails: {\tt  seokkimseok@gmail.com, 
seongmin0708@snu.ac.kr, \\
siyul.lee@kuleuven.be, pjy3873@snu.ac.kr}

\end{center}

\vspace{0.1cm}

\begin{abstract}

Microstates of BPS AdS black holes have been studied from the classical cohomologies 
of the maximal super-Yang-Mills theories, but their quantum natures have been conjectural. It was recently found that a classical black hole (fortuitous) cohomology 
in the $SO(7)$ theory is lifted by 1-loop corrections. We show that such 
lifts also happen in the $SU(2)$ theory, presenting both lifted/unlifted 
examples. In particular, the lightest fortuitous cohomology and its `hairy' versions are unlifted, 
while many heavier `core' fortuitous ones are lifted. 
We argue that the entropy of classical cohomologies in the Cardy limit 
is larger than the indicial entropy of strictly protected states 
by at least $\approx 1.2 \%$.

\end{abstract}

\end{titlepage}

\renewcommand{\thefootnote}{\arabic{footnote}}

\setcounter{footnote}{0}

\renewcommand{\baselinestretch}{1}

\tableofcontents

\renewcommand{\baselinestretch}{1.2}

\section{Introduction and discussions}

In the past few years, a program of exploring the microstates of BPS black holes in 
$AdS_5\times S^5$ has been studied extensively \cite{Chang:2022mjp,Choi:2022caq,Choi:2023znd,Chang:2023zqk,Budzik:2023vtr,
Budzik:2023xbr,Chang:2023ywj,Choi:2023vdm,Chang:2024zqi,deMelloKoch:2024pcs,Gadde:2025yoa,
Chang:2025mqp,Choi:2025bhi,Budzik:2025zvu}, based on the classical supercharge 
cohomologies for the $\frac{1}{16}$-BPS operators \cite{Grant:2008sk,Chang:2013fba}
in the $\mathcal{N}=4$ super-Yang-Mills theory.
While this setup provides a nice model for studying black holes and quantum gravity, the program is based on an assumption that 
the classical cohomologies would somehow represent the full quantum ones. 
This assumption is phrased more concretely by conjecturing that the spectrum 
of classical cohomologies, using the supercharge $Q$ of the classical interacting 
$\mathcal{N}=4$ Yang-Mills theory, would be identical to that of the 
$\frac{1}{16}$-BPS operators of the generic interacting quantum theory 
\cite{minwalla,Grant:2008sk}. Whether this conjecture is true or not, and if not 
how the quantum corrections modify the physics, have not been addressed until
recently.

Very recently, a counterexample to this conjecture was found in the $\mathcal{N}=4$ theory with $SO(7)$ gauge group \cite{Gadde:2025yoa,Chang:2025mqp,Choi:2025bhi}. 
A pair of classical cohomologies was shown to acquire anomalous dimensions 
after including 1-loop corrections to $Q$. 
This example was found while trying to test the S-duality between the 
$SO(7)$ and $Sp(3)$ theories with the conjecture assumed, finding an apparent 
mismatch of their spectra and then trying to resolve the tension with S-duality.

In this paper we study quantum corrections to the classical cohomologies 
in the simpler $SU(2)$ theory. We study 
classical $SU(2)$ cohomologies constructed in \cite{Choi:2023znd}, finding 
examples of both lifted and unlifted cohomologies at 1-loop.

To better understand these examples, we note that the cohomologies 
are classified into supergraviton type and the rest 
(which the black hole states should belong to, thus often morally called 
the `black hole' type). Algebraically, within 
the classical cohomologies, \cite{Chang:2024zqi} suggested
a closely related classification into the `monotone' and `fortuitous' cohomologies. 
The criterion is whether trace relations for finite $N$ matrices are 
used (fortuitous) or not (monotone) to ensure the $Q$-closedness of the operators. 
The supergravitons and the black hole states are respectively contained 
in the monotone and the fortuitous classes.

When classical cohomologies are lifted by 1-loop corrections, they become anomalous in 
superpartner pairs. The lifted pair in the $SO(7)$ theory mentioned 
above consists of a fortuitous and a monotonous one. This illustrates that 
the algebraic notions of the monotone/fortuity may become ambiguous at 
the quantum level. Also, this illustrates that the spectrum of multi-trace supergraviton 
type operators at finite $N$ (which had been believed to be well understood) 
defies a purely classical understanding, at least for the $SO(7)$ theory. 

We study several examples of classical cohomologies in the $SU(2)$ theory.
The `lightest' fortuitous cohomology which we call $O_0$ is unlifted at the 1-loop 
level. We further consider a few product cohomologies (which we dub `hairy black 
hole' states) of $O_0$ times gravitons, which are also unlifted at 1-loop. We then consider 
the heavier `core' fortuitous cohomologies constructed in \cite{Choi:2023znd}
that seem to not factorize into smaller cohomologies,
which we call $O_n$ with $n=1,2,\cdots$. 
We show that the three lightest ones $O_1$, $O_2$ and $O_3$ 
respectively pair with new classical
fortuitous cohomologies and get lifted at 1-loop.
The pairs that lift here consist of fortuitous cohomologies only.
If such quantum lifts are ubiquitous, they may alter the large charge BPS entropy. One may want to know 
by how much the classical cohomologies over-estimate 
the true $\frac{1}{16}$-BPS spectrum at large charges. 
We provide one such estimate by comparing the Cardy limit entropy of the index \cite{Romelsberger:2005eg,Kinney:2005ej}
and that of the classical cohomologies refined 
by a bonus $U(1)$ symmetry \cite{Chang:2013fba}. 
The over-estimation is nonzero but intriguingly small, around $\approx 1.2\%$.

Beyond the 1-loop effects on the classical cohomologies, 
further lifts are in principle possible at higher loops, at non-perturbative levels 
or at finite coupling. They have been recently discussed in the BMN matrix 
quantum mechanics \cite{Chang:2026ccg,Colin-Ellerin:2026eta}, where a 
non-renormalization theorem is established.

The remaining part of this paper is organized as follows. In Section \ref{sec:problem}, we 
explain the problem of and our computational strategies for assessing the 
quantum lifts of classical cohomologies. In Section \ref{sec:results}, we study several
$SU(2)$ classical cohomologies and discuss their 1-loop corrections.
Section \ref{sec:entropy} compares the Cardy entropy of the index with and without
the $U(1)$ refinement, and we discuss its implications.

\section{Quantum corrections of cohomologies}\label{sec:problem}

We start by establishing conventions for writing operators and acting the supercharge
in the $\mathcal{N}=4$ super-Yang-Mills theory.
Then we will introduce the problem of one-loop correction to supercharge cohomologies,
whose result will be presented in the next section.
We refer to \cite{Choi:2023znd} (see also \cite{Grant:2008sk,Chang:2013fba}) for more
detailed explanation for the setup of the cohomology problem.

Among 16 Poincar\'e and 16 conformal supercharges of the $\mathcal{N}=4$ super-Yang-Mills theory,
we choose $Q \equiv Q^4_-$ and its conjugate (within radial quantization) $S \equiv S^-_4 = Q^\dagger$.
In the free limit of the gauge theory, $\frac{1}{16}$-BPS local operators $O$ that preserve
the supercharges, i.e. $[Q, O \} = [Q^\dagger, O \} = 0$, consist of the following
elementary fields, also referred to as BPS letters.
Each of them transforms as an adjoint of the gauge group, $SU(2)$ throughout this paper.
\be\label{BPSfields}
\text{Scalars:}~ \phi^m~, \quad \text{chiralini:}~ \psi_m~, \quad
\text{gauge field:}~ f~, \quad \text{gaugini:}~ \lambda_\alpha~. \quad
(m = 1,2,3,~ \alpha = 1,2)~~
\ee
In addition, arbitrary derivatives (by $D_\alpha$) of each field at the origin $z^\alpha=0$
are leveraged to construct the full Hilbert space as local operators at the origin.
The indices signal the $\mathcal{N}=4$ superconformal algebra $PSU(2,2|4)$,
of which $PSU(1,2|3)$ commutes with the supercharge.
The BPS letters transform under the latter, thus $m$ is the (anti-)fundamental index
for the $SU(3)_R$ global symmetry, and $\alpha$ is the doublet index for
the $SU(2)$ Lorentz group (which often appears dotted in a broader context).
The gaugini are subject to the equation of motion $\epsilon^{\alpha \beta} D_\alpha \lambda_\beta = 0$. The ordinary derivatives $D_\alpha$ in the free theory will 
be promoted to the covariant derivatives in the interacting theory below.

When the coupling $g_{\rm YM}$ is turned on, the supercharge starts to act nontrivially
on the fields that were BPS at the free level. We are interested in the perturbative
regime of the Yang-Mills theory at weakly coupling,
so we expand the supercharge as
\be\label{fullQ}
Q = Q_0 + g_{\rm YM}^2 \, Q_1 + O(g_{\rm YM}^4)~.
\ee
The tree-level, or classical, $Q_0$ acts on the fields and their derivatives as
\begin{align}\label{Q0action}
[Q_0, \phi^m] &= 0~, & 
\{Q_0, \psi_m\} &= -i \epsilon_{mnl}[\phi^n, \phi^l]~, \nn \\
\{Q_0, \lambda_\alpha \} &= 0~, & 
[Q_0, f] &= i [\phi^m, \psi_m]~, \nn \\
&& [Q_0, D_{\alpha}] &= -i [\lambda_{\alpha}, ~~ \}~,
\end{align}
where (anti-)commutators treat each field as an adjoint matrix under the gauge group.

The action of the 1-loop level supercharge $Q_1$ was worked out in \cite{Budzik:2025zvu}.
It acts roughly as a double-derivative, meaning that it acts on two fields at a time,
as opposed to one as in the action of $Q_0$.
It is best written in terms of gauge components of the fields,
so let us expand each field as $\Phi = \Phi^A T^A$ with standard generators $T^A$
(which we take to be Pauli matrices for $SU(2)$), satisfying
\be
\tr{(T_A T_B)} = \kappa\delta_{AB}~, \quad
\tr{(T_A [T_B, T_C])}=\kappa f_{ABC}~.
\ee
Here, $\kappa$ is the dual Coxeter number of the gauge group, equal to $N = 2$
throughout this paper, and $f_{ABC}$ are the structure constants.
Then, the action of $Q_1$ on a pair of field components,
first where one of the pair does not contain any derivatives, is
(see (B.5-10) of \cite{Budzik:2025zvu}; overall factors are adapted to our convention of fields.)
\begin{align}
    Q_1 ( f^A \partial_1^m \partial_2^n \lambda^B ) 
        =& \frac{1}{4}\kappa^2 f_{ACD}f_{BCE} \frac{1}{m+n+2} \sum_{k=0}^m \sum_{l=0}^n \frac{1}{k+l+1} \binom{m}{k} \binom{n}{l} \nn \\
        & \cdot\Big[\partial_1^{k} \partial_2^{l} \partial_{\alpha} \lambda^D \,
        \partial_1^{m-k} \partial_2^{n-l} \partial^{\alpha} \lambda^E \Big]\, , \\
    Q_1 ( f^A \partial_1^m \partial_2^n f^B ) 
        =& \frac{1}{4}\kappa^2 f_{ACD}f_{BCE}\frac{1}{m+n+2} \sum_{k=0}^m \sum_{l=0}^n \frac{1}{k+l+1} \binom{m}{k} \binom{n}{l} \nn \\
        &\cdot\Big[ \partial_1^{k} \partial_2^{l} \partial_{\alpha} \lambda^D \,
        \partial_1^{m-k} \partial_2^{n-l} \partial^{\alpha} f^E
        - \partial_1^{k} \partial_2^{l} \partial_{\alpha} f^D \,
        \partial_1^{m-k} \partial_2^{n-l} \partial^{\alpha} \lambda^E \\
        &~~~ + \partial_1^{k} \partial_2^{l} \partial_{\alpha} (\psi_I)^D \,
        \partial_1^{m-k} \partial_2^{n-l} \partial^{\alpha} (\phi^I)^E
        - \partial_1^{k} \partial_2^{l} \partial_{\alpha} (\phi^I)^D \,
        \partial_1^{m-k} \partial_2^{n-l} \partial^{\alpha} (\psi_I)^E  \Big] \,, \nn \\
    \!\!\!\! Q_1 ( (\psi_I)^A \partial_1^m \partial_2^n (\phi^J)^B ) 
        =& \frac{1}{4}\kappa^2 \delta^J_I f_{ACD}f_{BCE} \frac{1}{m+n+2} \sum_{k=0}^m \sum_{l=0}^n \frac{1}{k+l+1} \binom{m}{k} \binom{n}{l} \nn \\
        & \cdot\Big[\partial_1^{k} \partial_2^{l} \partial_{\alpha} \lambda^D \,
        \partial_1^{m-k} \partial_2^{n-l} \partial^{\alpha} \lambda^E \Big]\,, \\
    \!\!\!\! Q_1 ( (\psi_I)^A \partial_1^m \partial_2^n (\psi_J)^B )  
        =& \frac{1}{2}\kappa^2 \varepsilon_{IJK} f_{ACD}f_{BCE} \frac{1}{m+n+2} \sum_{k=0}^m \sum_{l=0}^n \frac{1}{k+l+1} \binom{m}{k} \binom{n}{l} \nn \\
        & \cdot \Big[ \partial_1^{k} \partial_2^{l} \partial_{\alpha} \lambda^D \,
        \partial_1^{m-k} \partial_2^{n-l} \partial^{\alpha} (\phi^K)^E
        - \partial_1^{k} \partial_2^{l} \partial_{\alpha} (\phi^K)^D \,
        \partial_1^{m-k} \partial_2^{n-l} \partial^{\alpha} \lambda^E \Big]\, , \\
    Q_1 ( f^A \partial_1^m \partial_2^n (\phi^I)^B )
        =& \frac{1}{4}\kappa^2 f_{ACD}f_{BCE} \frac{1}{m+n+2} \sum_{k=0}^m \sum_{l=0}^n \frac{1}{k+l+1} \binom{m}{k} \binom{n}{l} \nn \\
        & \cdot\Big[\partial_1^{k} \partial_2^{l} \partial_{\alpha} \lambda^D \,
        \partial_1^{m-k} \partial_2^{n-l} \partial^{\alpha} (\phi^I)^E
        - \partial_1^{k} \partial_2^{l} \partial_{\alpha} (\phi^I)^D \,
        \partial_1^{m-k} \partial_2^{n-l} \partial^{\alpha} \lambda^E \Big] \, , \\
    Q_1 ( (\psi_I)^A \partial_1^m \partial_2^n f^B ) 
        =& \frac{1}{4}\kappa^2 f_{ACD}f_{BCE} \frac{1}{m+n+2} \sum_{k=0}^m \sum_{l=0}^n \frac{1}{k+l+1} \binom{m}{k} \binom{n}{l} \nonumber \\
        & \cdot\Big[\partial_1^{k} \partial_2^{l} \partial_{\alpha} (\psi_I)^D \,
        \partial_1^{m-k} \partial_2^{n-l} \partial^{\alpha} \lambda^E
        + \partial_1^{k} \partial_2^{l} \partial_{\alpha} \lambda^D \,
        \partial_1^{m-k} \partial_2^{n-l} \partial^{\alpha} (\psi_I)^E \nn \\
        &~~~ + 2\varepsilon_{IJK} \partial_1^{k} \partial_2^{l} \partial_{\alpha} (\phi^J)^D \,
        \partial_1^{m-k} \partial_2^{n-l} \partial^{\alpha} (\phi^K)^E  \Big] \, .
\end{align}
Here, $I,J,\cdots$ are $SU(3)_R$ indices and $A,B,\cdots$ are gauge adjoint indices,
and $\lambda$ without the $SU(2)$ index relates to our convention as $\lambda_\alpha = D_\alpha \lambda$.
Use of the unindexed auxiliary object $\lambda$ naturally reflects the equation of motion 
for $\lambda_\alpha$,
and underived $\lambda$ never appears in operators.
From these formulae, one can obtain the $Q_1$ action on a pair of fields where now both
may contain derivatives, using
\be
\partial_\alpha Q_1(f g) = Q_1((\partial_\alpha f) g) +Q_1(f (\partial_\alpha g))~,
\ee
recursively. Finally, on a generic operator (which must consist of at least 2 fields) it acts
as a generalization of the Leibniz rule:
\be
Q_1 (f_1 \cdots f_n) = \sum_{i<j} (-1)^\# Q_1 (f_i f_j) f_1 \cdots f_{i-1} f_{i+1} \cdots f_{j-1} f_{j+1} \cdots f_n~,
\ee
where the $(-1)^\#$ factor encodes the fermion anti-commutation sign that may
arise from bringing $f_i$ and $f_j$ to the front.

We note that the action of $Q_0$ always increases the number of fields by one.
It is in this sense that the number of fields $Y$ is a well defined
quantum number at the classical level
(also known as the bonus $U(1)$ symmetry \cite{Chang:2013fba}),
under which $Q_0$ is charged by unity.
This bonus symmetry will play the key role in further refining the index in Section \ref{sec:entropy}.
On the other hand, $Q_1$ has zero charge under this symmetry,
breaking the symmetry at the quantum level.
In general, $Q_n$ can be considered to carry charge $1-n$ in this symmetry \cite{Budzik:2023xbr}.

On a related note, the full quantum supercharge is originally
\be
Q = Q_{\rm free} + g_{\rm YM} \, Q_0 + g_{\rm YM}^2 \, Q_1 + O(g_{\rm YM}^3)~.
\ee
The free supercharge $Q_{\rm free}$ annihilates all BPS fields \eqref{BPSfields},
thus it is practical to ignore it entirely.
The next level $Q_0$, often called the classical supercharge,
actually acts at the ``half-loop'' $\sim g_{\rm YM}^1$ level,
so that the BPS hamiltonian $H = \{ Q, Q^\dagger \} \sim g_{\rm YM}^2$ is one-loop.
For our convenience, we perform a global field redefinition $\Phi \to g_{\rm YM}^{-1} \Phi$.
As a result, the classical supercharge $Q_0$ that lived in the half-loop level $\sim g_{\rm YM}^1$
becomes a genuine tree-level action $\sim g_{\rm YM}^0$,
and all higher $Q_n$ settle at $n$-loop levels $\sim g_{\rm YM}^{2n}$.
Hereafter we take the field-redefined convention \eqref{fullQ}.

The recent black hole microstate program has mainly focused on 
the classical cohomologies,
i.e. operators that are annihilated by $Q_0$ with equivalent relations spanned by $Q_0$-exact operators.
Many examples of classical cohomologies were found.
We limit to the $SU(2)$ gauge group.
First there are super-graviton cohomologies, which consist of the single-trace operators,
\begin{align}\label{single-grav}
  & \tr{(\phi^{m}\phi^{n})}~, ~~~
  \tr{(\phi^m \lambda_\alpha)}~, ~~~ 
  \tr{(\lambda_{+} \lambda_{-})}~, ~~~
  \tr{(\phi^m \psi_{n})}-{\textstyle \frac{1}{3}}\delta^m_n \tr{(\phi^l \psi_l)}~, \\
  &
  \tr{(\lambda_\alpha \psi_{m} - \epsilon_{mnp} \phi^n D_{\alpha}\phi^p)}~, ~~~
  \tr{(\phi^m f -{\textstyle\frac{1}{4}}\epsilon^{mnp} \psi_n \psi_p)}~, ~~~
  \tr{(\lambda_\alpha f - {\textstyle\frac{2}{3}} \psi_m D_\alpha \phi^m + {\textstyle\frac{1}{3}}\phi^m D_\alpha \psi_m)}~, \nn
\end{align}
their descendants and their products.
Then there are black hole type, or fortuitous, cohomologies, that are not gravitons.
An infinite tower ($n = 0,1,2, \cdots$) of such cohomologies, that are fermionic and singlet in the global symmetry $SU(3)_R$,
were constructed in \cite{Choi:2022caq,Choi:2023znd}:
\begin{eqnarray}\label{On-summary}
  O_n &=& \tr{(f f)}^n \epsilon^{c_1c_2c_3} \tr{(\phi^a \psi_{c_1})}
  \tr{(\phi^b \psi_{c_2})} \tr{(\psi_a \psi_b \psi_{c_3})} \nn \\
  &&- \frac{n}{2} \tr{(f f)}^{n-1}\epsilon^{b_1b_2b_3}\epsilon^{c_1c_2c_3} \tr{(f \psi_{b_1})}
  \tr{(\phi^a \psi_{c_1})} \tr{(\psi_{b_2} \psi_{c_2})} \tr{(\psi_a \psi_{b_3} \psi_{c_3})} \\
  &&+ \frac{2n^2 +n}{432}
  \tr{(f f)}^{n-1}\epsilon^{a_1a_2a_3}\epsilon^{b_1b_2b_3}\epsilon^{c_1c_2c_3}
  \tr{(\psi_{a_1} \psi_{b_1} \psi_{c_1})}
  \tr{(\psi_{a_2} \psi_{b_2} \psi_{c_2})}
  \tr{(\psi_{a_3} \psi_{b_3} \psi_{c_3})}~. \nn
\end{eqnarray}

We are ultimately interested in cohomologies with respect to the full supercharge $Q$, \eqref{fullQ}.
In this paper, we shall study the cohomologies in the perturbation theory in $g_{\rm YM}$.
For this, let us also expand the cohomology $\CO$ in powers of $g_{\rm YM}^2$,
(we use the caligraphic $\CO$ and $\CO_n$ to denote generic operators in this paragraph,
not to be confused with ordinary $O_n$ \eqref{On-summary} which refers to the specific
$SU(2)$ cohomologies)
\be\label{fullO}
\CO = \CO_0 + g_{\rm YM}^2 \, \CO_1 + O(g_{\rm YM}^4)~.
\ee
Then,
\begin{eqnarray}\label{QOpert}
Q \, \CO
&=& (Q_0 + g_{\rm YM}^2 \, Q_1) (\CO_0 + g_{\rm YM}^2 \, \CO_1) + O(g_{\rm YM}^4) \nn\\
&=& Q_0 \CO_0 + g_{\rm YM}^2 \, (Q_1 \CO_0 + Q_0 \CO_1) + O(g_{\rm YM}^4)~.
\end{eqnarray}
Thus, for $\CO$ to be a full quantum cohomology, we require at least two conditions
from the classical and the 1-loop order respectively:
\begin{enumerate}
\item we must start with a classical cohomology $\CO_0$ with respect to $Q_0$, and
\item $Q_1 \CO_0$ must be $Q_0$-exact.
\end{enumerate}
The first condition is trivial but meaningful, because it states that the classical cohomologies
found in various theories so far are the first steps towards
the quantum cohomologies.
Starting with an established classical cohomology $\CO_0$ as required by 1, we now turn to 2.
If 2 holds, we may let $Q_1 \CO_0 = Q_0 (-\CO_1)$.
Then $\CO_0$ with the 1-loop correction $\CO_1$ is closed under $Q$ until the 1-loop level.
Of course, it remains to check all higher orders to argue that $\CO_0$ can be 
(perturbatively) completed to a full quantum cohomology $\CO$.
If, however, 2 does not hold, then we immediately know that $\CO_0$ acquires an anomalous dimension
at the 1-loop level and thus cannot be completed into a full quantum cohomology.
In conclusion, a classical cohomology $\CO_0$ may lift because it is not $Q$-closed at
the 1-loop level, when $Q_1 \CO_0$ is not $Q_0$-exact.

When a classical cohomology $\CO_0$ lifts at the quantum level by $Q_1\mathcal{O}_0$ 
failing to be $Q_0$-exact,
it must pair with another classical cohomology that also lifts.
Since $\CO$ and $Q \CO$ are two operators paired up by the supercharge,
the classical cohomology that pairs with $\CO_0$ to lift together is $Q_1 \CO_0 + Q_0 \CO_1$
or equivalently (cohomologously) $Q_1 \CO_0$.
It is easy to show that $Q_1 \CO_0$ is indeed a classical cohomology.
It follows from nilpotency of $Q$ that $\{Q_0, Q_1 \} = 0$. Then,
\be
Q_0 (Q_1 \CO_0) = - Q_1 Q_0 \CO_0 = 0,
\ee
i.e. $Q_1 \CO_0$ is $Q_0$-closed,
where the last equality holds because $\CO_0$ is a classical cohomology.
Moreover, $Q_1 \CO_0$ is not $Q_0$-exact by assumption.
Note that
\be
Q_1 (Q_1 \CO_0) = -\{Q_0, Q_2\} \CO_0 = - Q_0 (Q_2 \CO_0)~,
\ee
is $Q_0$-exact, meaning that the classical cohomology $Q_1 \CO_0$ does not lift
because it is not $Q$-closed as in the previous paragraph.
However, it lifts because it becomes $Q$-exact due to quantum corrections.
More concretely, given a classical cohomology $\CO_0$,
it is lifted because it is $Q$-exact at 1-loop if and only if
\be\label{1-loop-Q-exact}
\CO_0 = Q_1 \L_0 + Q_0 \L_1~,
\ee
for some classical cohomology $\L_0$ and any operator $\L_1$.

Therefore, a classical cohomology $\CO_0$ can lift at the one-loop level through two mechanisms;
by $Q_1 \CO_0$ not being $Q_0$-exact, or by being ($Q_0$-cohomologous to)
a $Q_1$-image of a classical cohomology.
Two types of lifted cohomologies are matched one-to-one.

As we will exploit in the next section, there are special cases where the second mechanism is precluded.
We can define a particular combination of the quantum numbers
($R_{1,2,3}$ the R-charges, $J_{1,2}$ the angular momenta and the number of fields $Y$,
we refer to Table 1 of \cite{Choi:2023znd} for the full charge contents)
\be
d_{\rm BMN} \equiv R_1+R_2+R_3 + \frac{J_1 + J_2}{2} - Y~.
\ee
$\phi^m$, $\psi_m$ and $f$ all carry zero of this quantum number,
while each $\lambda_\alpha$ and $D_\alpha$ carries $\frac12$.
Importantly, $Q_0$ carries $0$ while $Q_1$ carries $1$.
Therefore, classical cohomologies are naturally graded by this number,
and especially the subspace of classical cohomologies where $d_{\rm BMN} = 0$,
called the BMN subsector, has been studied exclusively
\cite{Choi:2023vdm,deMelloKoch:2024pcs,Gaikwad:2025ugk}.
$d_{\rm BMN}$ can be thought of as a measure of the distance from the BMN subsector,
and the action of $Q_1$ monotonously increases it.
Since $d_{\rm BMN} \geq 0$ for any local operator, if a classical cohomology $\CO_0$
belongs to the BMN subsector, i.e. $d_{\rm BMN}(\CO_0) = 0$,
it cannot be $Q_1$-exact and the second mechanism is not viable.

In the next section we will act $Q_1$ on various established classical cohomologies
and investigate their $Q_0$-exactness, to determine if they lift at the 1-loop level
under the first mechanism.
In most of discussion, the classical cohomologies belong to the BMN subsector,
so this suffices to determine whether they lift at the 1-loop level.

Determining $Q_0$-exactness of an operator is in general a daunting task.
It has been the bottleneck in many works on classical cohomologies \cite{Chang:2022mjp,Choi:2023znd,Choi:2023vdm},
and it was skipped using certain tricks in \cite{deMelloKoch:2024pcs}.
In this work we take no shortcut, but proceed in the most honest and brutal way
of constructing the complete $Q_0$-exact basis in that charge sector,
and investigating whether a given target operator can be expressed as their
linear combination.
In the rest of this section, we explain on general grounds three technical details involved
in this process. We hope that their application to specific contexts in the next section
will make them more clear.

The first is about how to build the basis.
The question is if we can write $Q_1 \CO_0 = Q_0 \L$ with some $\L$,
so we'd like to construct a basis that spans the space $\L$ lives in.
Since $Q_0$ and $Q_1$ carry the same set of charges, namely two angular momenta
and three R-charges, $\L$ must carry the same set of charges as $\CO_0$.
Moreover, we have observed that $Q_0$ and $Q_1$ are charged $1$ and $0$ respectively
under the number of fields. Although the latter is not an exact symmetry of the theory,
it is homogeneous per perturbative orders in $g_{\rm YM}$, so
we can practically restrict the number of fields appearing in $\L$ to be one fewer than $\CO_0$.

Actually, one can impose a stronger constraint on the basis operators for $\L$.
If $Q_1 \CO_0$ transforms as a certain representation under the global symmetry group $SU(3)_R$,
$\L$ must also be in the same representation, since $Q_0$ commutes with the global symmetry.
For example, when $\L$ is expected to be a singlet, we can restrict the basis operators
to have all their $SU(3)_R$ index covariantly contracted.
On the other hand, all `monomials' that appear as the summand in the implicit summation
for the contraction, would all have the correct charges.
Therefore, mathcing the representation is a stronger requirement than matching only the
Cartans of the symmetry group, resulting in a basis with a smaller dimension.
There is also a downside to this approach,
namely that basis operators of definite representations involving contractions,
are in general sums over many monomial-like basis operators.
So the operators are longer when expanded inside a computer
than the monomial-like operators that only match the Cartans.
Balancing between complexity of each operator and size of the basis,
we take both methods and cross-check the result for most applications.
One could also impose the same covariance principle with the $SU(2)$ Lorentz group indices,
which we do not do.

The second detail is about turning the algebraic problem into a linear algebra problem.
Once the basis for $\L$, and thus the basis of $Q_0$-exact operators is built,
it becomes an algebraic problem to determine
linear independence between them and the target operator $Q_1 \CO_0$.
An obvious way to proceed is to turn it into a linear algebra problem by
expanding the operators into polynomials and extracting their coefficient matrix.
In the coefficient matrix, each column represents a basis operator,
and each row represents a monomial that appears in some operator.
Concretely, let $\{ m_1 , \cdots , m_a \}$ be the union of all monomials that appear
in polynomials $\{ p_1, \cdots , p_b \}$ as well as the target operator $Q_1 \CO_0$.
The matrix will have dimensions $a \times b$:
\be
A = \begin{bmatrix}
\text{coef. of } m_1 \text{ in } p_1 & \cdots & \text{coef. of } m_1 \text{ in } p_b \\
\vdots & \ddots & \vdots \\
\text{coef. of } m_a \text{ in } p_1 & \cdots & \text{coef. of } m_a \text{ in } p_b
\end{bmatrix}~.
\ee
In this way, each basis operator is effectively represented by a numerical column vector.
$Q_1 \CO_0$ is similarly expressed as a column vector $\vec{b}$.
Whether $Q_1 \CO_0$ is $Q_0$-exact, is now equivalent to whether $\vec{b}$ is in the
column space of $A$, in other words whether
\be\label{Ax=b}
A \cdot \vec{x} = \vec{b}~~\text{has a solution}~\vec{x}
\qquad \Leftrightarrow \qquad
{\rm rank}\,(A) = {\rm rank}\,(A; b)~,
\ee
where $(A; b)$ refers to the augmented matrix.

One practical problem of this approach is that often there are far more monomials (rows)
than there are polynomials (columns), resulting in a very rectangular coefficient matrix $a \gg b$.
It is inefficient in a sense that we are using too many ($a$) datasets to represent a column space
whose dimension is at most $b$: linear dependency between $b$ columns can
in principle be encoded in as few as $b$ rows.

A numerics-assisted approach has been developed in \cite{Choi:2023vdm} against such an inefficiency.
In this approach, one substitutes the field components (collectively called $\Phi$), 
or variables of the polynomials,
with $c \geq b$ sets of random integers, to build a $c \times b$ matrix:
\be
A_{\rm rand} = \begin{bmatrix}
\left. p_1 \right|_{\Phi \to \text{set }1} & \cdots & \left. p_b \right|_{\Phi \to \text{set }1} \\
\vdots & \ddots & \vdots \\
\left. p_1 \right|_{\Phi \to \text{set }c} & \cdots & \left. p_b \right|_{\Phi \to \text{set }c}
\end{bmatrix}~.
\ee
In this way, each basis operator is represented by a numerical column vector of length $c$.
$Q_1 \CO_0$ is similarly expressed as a vector $\vec{b}_{\rm rand}$.
If there is a relation between polynomials, including the target operator,
the relation must also hold between the numbers obtained by substituting with the same set of numbers,
so the relation carries on to that between the vectors.
Conversely, a relation between the vectors do not guarantee corresponding relation
between the polynomials, because it is in principle possible that a non-zero polynomial
miraculously vanishes for some choice of random numbers.
Of course, chance for this error can be astronomically reduced by introducing a margin
$c-b$ between the number of rows and columns.\footnote{$c$ only needs to be
equal or greater than the rank $r \leq b$ of the matrix.
$c \geq b$ is the safest choice before knowing the rank.}
This method allows us to turn the algebraic problem into the linear algebra problem \eqref{Ax=b}
but with a nearly square matrix $A_{\rm rand}$, which is often far smaller than
the full coefficient matrix $A$.
Again, we take both methods and cross-check the result.

Finally, for \eqref{Ax=b} it remains to compute the ranks or perform a Gaussian elimination
with a (possibly very large) matrix of exact rational numbers, either $A$ or $A_{\rm rand}$.
A method often employed in mathematics or computer sciences for this type of linear problem
is to work on a finite field $\mathbb{Z}_p$ with a large prime number $p$,
by converting every number in the input and during every step of the process into an
integer from $0$ to $p-1$, boosting the arithmetics.
An error may occur e.g. when a result turns out to be miraculously a multiple of $p$,
but chance for such errors can be exponentially reduced by repeating computations
over multiple values of $p$, and using as large $p$ as permitted.
We take the following 4 values of $p$ in our computations, which are the standard choice
as they are the 4 largest prime numbers that fit in \texttt{int32} data format:
\begin{eqnarray}
&&2^{31} - 1 = 2147483647\ \ \ ,\ \qquad   
2^{31} - 19 = 2147483629\ ,\nonumber\\ 
&&2^{31} - 61 = 2147483587\ \ ,\ \qquad
2^{31} - 69 = 2147483579\ .
\end{eqnarray}
We refer to Appendix \ref{app:modp} for further and more rigorous mathematical context on this approach.

\section{Results}\label{sec:results}

In this section, we report whether each of the classical cohomologies in the
$\mathcal{N}=4$ super-Yang-Mills with $SU(2)$ gauge group is lifted at the 1-loop level.
As we have explained in the previous section, there are two mechanisms of the lift.
The first happens when $Q_1 \CO_0$ is not $Q_0$-exact (failing to be $Q$-closed 
with the quantum supercharge), and the second happens when $\CO_0 = Q_1 \L_0 + Q_0 \L_1$
for some classical cohomology $\L_0$ and an operator $\L_1$ (becoming $Q$-exact 
after including the quantum effects to $Q$). We discuss both possibilities.

\subsection{Threshold fortuitous cohomology $O_0$}\label{sec:O0}

We first consider the threshold fortuitous cohomology found in \cite{Chang:2022mjp,Choi:2022caq}.
\be\label{theO0}
O_0 = \epsilon^{c_1c_2c_3} \tr{(\phi^a \, \psi_{c_1})}
  \tr{(\phi^b \, \psi_{c_2})} \tr{(\psi_a \, \psi_b \, \psi_{c_3})}~.
\ee
It carries charges $(R_1, R_2, R_3, J_1, J_2) = (\frac32, \frac32, \frac32, \frac52, \frac52)$,
often summarized into $j \equiv 2(R_1+R_2+R_3) + 3(J_1+J_2) = 24$.
The number of fields is  $Y = 7$,
and $d_{\rm BMN} \equiv R_1+R_2+R_3 + \frac{J_1 + J_2}{2} - Y = 0$, i.e. it belongs to the BMN subsector.
Evaluating $Q_1 O_0$ gives
\begin{align}
        Q_1 O_0 =&~ 8\Tr(\phi^{i}\,\psi_{j})\tr(\lambda^{\alpha}\,\psi_{(i})\tr(\psi_{k)}\,(D_{\alpha}\phi)^{j}\,\phi^{k}) \nn\\
        &+ 4 \tr(\phi^{i}\,\psi_{[j})\,\tr(\phi^{k}\,\psi_{k]})\,\tr(\psi_{i}\,(D^{\alpha}\phi)^{j}\,\lambda_{\alpha}) \nn\\ 
        &+ \epsilon^{abc}\tr(\phi^{i}\,\psi_{a})\, \tr(\lambda^{\alpha}\,\psi_{b})\tr(\lambda_{\alpha}\,\psi_{i}\,\psi_{c})\,,
\end{align}
up to an overall factor. This carries charges $(2,2,2,2,2)$ and, like $O_0$,
is an $SU(3)_R$ singlet and has $Y = 7$.

We now investigate whether $Q_1 O_0 = Q_0 \L$ for some $\L$.
Since $Q_0$ carries the same set of charges as $Q_1$ but raises $Y$ by one,
the pre-image $\Lambda$ must carry the same charges as $O_0$,
i.e. $(\tfrac32,\tfrac32,\tfrac32,\tfrac52,\tfrac52)$ and consist of $Y=6$ letters,
so that $Q_0\Lambda$ may land back on $Q_1 O_0$.
In addition, $\L$ must be $SU(3)_R$ singlet, since $Q_0$ commutes with the global symmetry.
For $SU(2)$ gauge invariance, the six letters must arrange into three two-letter traces.
As we have explained in Section \ref{sec:problem}, the basis for the space of $\L$-candidates
can be constructed in two different ways.
\begin{enumerate}
    \item \underline{Matching the charges.} We enumerate all gauge invariant operators with charges $(\tfrac32,\tfrac32,\tfrac32,\tfrac52,\tfrac52)$ and $Y=6$.
    
    \item \underline{Matching the representation.} In addition to matching the charges, we require
    the basis operators to be $SU(3)_R$-singlets, just like $Q_1O_0$ itself.
    Abstractly, we are projecting the space of 1. onto singlets.
    Practically, we covariantly contract all $SU(3)_R$ indices.
\end{enumerate}
Given the space of $\L$-candidates from either method, we act $Q_0$ on them to span
the space of $Q_0$-exact operators in the charge sector of $Q_1O_0$.
From the first method, the resulting $Q_0$-exact space has rank $1903$.
From the second method, we obtain a space of rank $149$, which is a subset of the previous one.

We showed explicitly that $Q_1 O_0$ lies in both $Q_0$-exact spaces.
Moreover, $O_0$ lives in the BMN subsector, so it cannot be the $Q_1$-image of another cohomology.
Therefore, the $SU(2)$ threshold fortuitous cohomology $O_0$ \eqref{theO0} is
\emph{not} lifted at one loop.

This result is much expected.
If $O_0$ lifts, there must be another classical cohomology with same $j = 24$ and $Y = 7$
as $O_0$ to pair with.
However, a full scan over classical cohomologies till $j=25$ in the $SU(2)$ gauge theory
has been carried out in \cite{Chang:2022mjp},
which found that there is exactly one classical cohomology with $j=24$ and $Y=7$, namely $O_0$.
In particular, no graviton can have an odd number of fields in the $SU(2)$ gauge theory
given the single trace structure \eqref{single-grav},
so the pairing between a fortuitous and a graviton cohomologies as occured in the $SO(7)$ theory
\cite{Choi:2025bhi} does not happen for the threshold $O_0$, nor for any $O_n$ as we shall shortly see.

\subsection{Hairy black hole cohomologies of $O_0$}\label{sec:hairy}

Next, we investigate three types of hairy black hole cohomologies found in
\cite{Choi:2023znd}, obtained as products of some graviton cohomology and 
the threshold cohomology $O_0$ \eqref{theO0}. It will turn out that they are 
all unlifted at the 1-loop level.

The first example is the product between a graviton
$(w_2)^m = \tr{(\phi^m f -{\textstyle\frac{1}{4}}\epsilon^{mnp} \psi_n \psi_p)}$
that appeared as one of \eqref{single-grav}, and $O_0$.
This product was shown to be another classical cohomology,
and it is the lightest (in terms of the charge $j$) such product between a graviton and $O_0$,
surviving the partial no-hair theorem \cite{Choi:2023znd}.
Dressing of black hole cohomologies by this particular graviton is speculated
to be universal, even in the $SU(3)$ gauge theory \cite{Choi:2023vdm,deMelloKoch:2024pcs}.
Without loss of generality, we consider the top component of the fundamental representation under $SU(3)_R$:
\begin{eqnarray}\label{w2O0}
    w_2 O_0 = \tr{\left(\phi^1f-\frac{1}{2}\psi^2\psi^3\right)} \cdot
    \epsilon^{c_1c_2c_3} \tr{(\phi^a \, \psi_{c_1})} \tr{(\phi^b \, \psi_{c_2})}
    \tr{(\psi_a \, \psi_b \, \psi_{c_3})}\,.
\end{eqnarray}
It carries charges $(R_1, R_2, R_3, J_1, J_2) = (\frac52, \frac32, \frac32, \frac72, \frac72)$,
$j = 32$, $Y = 9$ and thus $d_{\rm BMN} = 0$.

To investigate whether $Q_1 (w_2 O_0) = Q_0 \L$ for some $\L$,
we construct the basis for the space of $\L$-candidates.
$\L$ must carry the same charges $(R_1, R_2, R_3, J_1, J_2)$ as $w_2O_0$, and $Y = 8$.
Again, we can take either of the two approaches in constructing this space:
matching only the charges or in addition the $SU(3)_R$ representation $[1,0]$.
We choose the first approach. Acting $Q_0$ to span the space of $Q_0$-exact operators
appropriate for $Q_1 (w_2O_0)$, we find a space of rank $14610$.

We proved that $Q_1 (w_2O_0)$ lies within this $Q_0$-exact space.
Moreover, $w_2 O_0$ lives in the BMN subsector, so it cannot be the $Q_1$-image of another cohomology.
Therefore, the lightest hairy black hole cohomology \eqref{w2O0} is \emph{not} lifted at one loop.

There are many more hairy black hole cohomologies of this type, either 
constructed explicitly or guessed from patterns of the index. 
We would like to discuss two more such cohomologies which dress $O_0$ 
(or its superconformal descendant), at charges $j=36$. Our original motivation 
for studying these was related to their possible quantum mixing/pairing
with the core fortuitous cohomology $O_1$, which will be discussed in Section
\ref{sec:On}. Here we present our results on these hairy cohomologies, postponing 
the discussion related to $O_1$ to Section \ref{sec:On}.

The next hairy cohomology we would like to discuss is 
\begin{equation}\label{O1prime}
  O_1^\prime \equiv \epsilon_{mnp}(Q_+^mQ_+^nO_0)(w_2)^p
\end{equation}
where $Q_+^m$ are some of the classical supercharges in $SU(2|3)\subset PSU(2,2|4)$ 
that acts within the BMN sector, and $(w_2)^p$ are the graviton cohomologies 
introduced above. That is, these are the same gravitons dressing the 
superconformal descendants of $O_0$ rather than $O_0$ itself. 
This was shown to be not $Q_0$-exact \cite{Choi:2023znd}, representing a classical cohomology.
This cohomology has charges $(R_1, R_2, R_3, J_1, J_2) = (\frac32, \frac32, \frac32, \frac92, \frac92)$,
$Y=9$ and $d_{\rm BMN} = 0$.

Our last example of hairy black hole cohomology is
\be\label{O1pp}
O_1^{\prime\prime} = \frac18 \,
\partial^{\alpha} \tr{\big( f\lambda_{\alpha}-\frac{2}{3}D_{\alpha}\phi^n\psi_n+\frac{1}{3}\phi^nD_{\alpha}\psi_n \big)} \cdot O_0~.
\ee
\cite{Choi:2023znd} conjectured the existence of this classical cohomology
(that is, not $Q_0$-exact) related to the physics of a core cohomology $O_1$ 
that we will discuss in Section \ref{sec:On}.
This has charges $(2,2,2,4,4)$, $Y=9$ and $d_{\rm BMN} = 1$.

We ask if these two, namely $O_1^\prime$ and $O_1^{\prime\prime}$, are lifted at one loop.
Recall from Section \ref{sec:problem} that there are two mechanisms through which a
classical cohomology may lift: it may stop being $Q$-closed, or become $Q$-exact at one loop.
Note that the second mechanism is automatically ruled out for $O_1^\prime$, but not for $O_1^{\prime\prime}$,
because the latter is not in the BMN subsector.
We now show that both do \emph{not} lift through either mechanism.

First, for the fate of $O_1^\prime$, we need to decide if $Q_1 O_1^\prime$ is $Q_0$-exact,
i.e. is $Q_0 \L$ for some operator.
We proceed with more or less same computations as in Section \ref{sec:O0}.
The pre-image basis (for $\L$) carries the same charges as $O_1^\prime$ but $Y=8$,
and is distributed into $4$ two-letter traces.
We build the basis using both methods for a cross-check;
matching the charges and matching the full $SU(3)_R$ representation (which is singlet).
The $Q_0$-exact space obtained by acting $Q_0$ on both bases have dimensions $15089$ and $912$ respectively.
$Q_1 O_1^\prime$ is included in both spaces, so it is $Q_0$-exact.
Since $O_1^\prime$ is in the BMN subsector,
we conclude that $O_1^\prime$ is not lifted at one loop.

Then for $O_1^{\prime\prime}$, we need to show against both mechanisms.
For the first, we have constructed the basis for the $Q_0$-exact space
with charges corresponding to $Q_1 O_1^{\prime\prime}$, restricted to the $SU(3)_R$ singlets.
The singlet $Q_0$-exact space has dimension $15096$,
and $Q_1 O_1^{\prime\prime}$ is included in this space, so it is $Q_0$-exact.

For the second mechanism, i.e. whether $O_1^{\prime\prime} = Q_1 \L_0 + Q_0 \L_1$,
we need to construct both bases for $\L_0$ and $\L_1$.
Since $O_1^{\prime\prime}$ lives in the same charge sector as $Q_1 O_1^\prime$,
we can recycle the $912$-dimensional space of $SU(3)_R$ singlets just introduced.
On the other hand, $\L_0$ has charges $(\frac32, \frac32, \frac32, \frac92, \frac92)$ and $Y=9$.
We construct the basis for such operators that are singlets, and act $Q_1$ to obtain a
25-dimensional space of singlet $Q_1 \L_0$.\footnote{Here, we construct the basis
for all possible singlet operators $\L_0$, whereas $\L_0$ needs to be $Q_0$-closed.
However, since all $Q_0$-closed operators are included in our basis,
this will not affect the conclusion.}
Combining with the 912-dimensional space of $Q_0 \L_1$,
we obtain a 930-dimensional space, accounting for 7 dimensions of overlap.
$O_1^{\prime\prime}$ is not included in this space, so it is not lifted at the 1-loop level by
becoming $Q$-exact.

We note that, this implies that $O_1^{\prime\prime}$ is itself not $Q_0$-exact.
It is also easy to show that it is $Q_0$-closed, so we conclude that 
$O_1^{\prime\prime}$ is a classical cohomology, as conjectured in \cite{Choi:2023znd}.

To conclude, we showed that three hairy black hole cohomologies that dress $O_0$,
namely \eqref{w2O0}, \eqref{O1prime} and \eqref{O1pp},
are not lifted at the one-loop level. It will be interesting to know if this 
is linked to the protection of $O_0$ itself, and if they may be protected to all (perturbative) orders.
(See Section \ref{sec:On} for related discussions.)

\subsection{Heavier core fortuitous cohomologies}\label{sec:On}

We now consider the next entries of the infinite tower of fortuitous cohomologies constructed in \cite{Choi:2023znd}. Explicit expression for the cohomologies $O_n$ labeled by an integer $n \geq 0$ was presented in \eqref{On-summary}, of which the $n=0$ case is the threshold cohomology $O_0$ of Section~\ref{sec:O0}. They carry charges $(R_1, R_2, R_3, J_1, J_2) = (\frac32, \, \frac32, \, \frac32, \, \frac52+2n, \, \frac52+2n)$, corresponding to $j \equiv 2(R_1+R_2+R_3) + 3(J_1+J_2) = 24 + 12n$.
The number of fields is $Y = 7+2n$, and it follows that $d_{\rm BMN} = 0$.

In contrast to $O_0$, we find that the $O_n$ cohomologies at $n=1,2,3$ are \emph{lifted} at one loop.
To test whether $Q_1 O_n$ is $Q_0$-exact, we proceed with same computations as in Section~\ref{sec:O0}, with appropriate charge sectors. The pre-image basis (for $\L$) carries the same charges as $O_n$
and consists of $Y=6+2n$ letters distributed into $3+n$ two-letter traces.
For $O_1$ and $O_2$, we build the bases using both methods for a cross-check.
For $O_3$, the first method results in a space too large to handle,
so we only take the second method of matching the $SU(3)_R$ representation which is a singlet.
\begin{itemize}
    \item $O_1$.
    This $Q_0$-exact space has been built in the previous subsection.
    The space only matching the charges has rank $15089$,
    and the $SU(3)_R$-singlet subspace has rank $912$.
    $Q_1 O_1$ is absent from the space of $Q_0$-exact operators.
    \item $O_2$. The corresponding ranks are $27468$ for the full charge sector and $1737$ for the singlet space. $Q_1 O_2$ is absent from this space.
    \item $O_3$. We only spanned the $SU(3)_R$-singlet space, which has rank $2043$.
    $Q_1 O_3$ is absent from this space.
\end{itemize}
We conclude that all of $O_1$, $O_2$ and $O_3$ are lifted at the one-loop level.

Even though these three cohomologies are lifted, they cannot pair with graviton cohomologies.
As we mentioned towards the end of Section \ref{sec:O0}, graviton cohomologies in the $SU(2)$ theory
must have even number of fields. Since $Q_1 O_n$ carries an odd number of fields, namely $Y = 7+2n$,
the classical cohomologies that pair with $O_n$ must be fortuitous.
This is in contrast to the pairing between a fortuitous and a graviton cohomologies in
the $SO(7)$ theory \cite{Choi:2025bhi}.

Additionally, we investigate the quantum superparter $Q_1O_1$ of $O_1$, 
as their pairing/lifting pattern reveals intriguing features. 
As $Q_1 O_1$ is $Q_0$-closed but not $Q_0$-exact, it represents a nontrivial classical cohomology class.
Furthermore, we find that this cohomology splits into two non-trivial inequivalent $Q_0$ cohomologies,
\begin{equation}
Q_1 O_1 = O_1^{\prime\prime}+ \widetilde{O}_1\ ,
\end{equation}
where $O_1^{\prime\prime}$ is defined in \eqref{O1pp} and $\widetilde{O}_1$ is given by
{\allowdisplaybreaks
    \hspace*{-1cm}
\begin{align}
        240 \, \widetilde{O}_1 ~=~~
        &480\,\tr(f\,f)\,\tr(\phi^{i}\,\psi_{j})\,\tr(\lambda^{\alpha}\,\psi_{(i})
        \,\tr(\psi_{k)}\,D_{\alpha}\phi^{j}\,\phi^{k})\nonumber \\ 
        &+240\,\tr(f\,f)\,\tr(\phi^{i}\,\psi_{[j})\,\tr(\phi^{k}\,\psi_{k]})\,\tr(\psi_{i}\,D^{\alpha}\phi^{j}\,\lambda_{\alpha})\nonumber \\ 
        &+60\,\epsilon^{a_1 a_2 a_3}\,\tr(f\,f)\,\tr(\phi^{i}\,\psi_{a_1})\, \tr(\lambda^{\alpha}\,\psi_{a_2})\,\tr(\lambda_{\alpha}\,\psi_{i}\,\psi_{a_3})\nonumber \\          
        &-480\,\tr(\phi^{a}\,\psi_{c})\,\tr(\phi^{b}\,\psi_{d})\,\tr(D^{\alpha}\phi^{d}\,\psi_{(a})\,\tr(\psi_{b)}\,f\,D_{\alpha}\phi^{c})\nonumber \\
        &-15\epsilon^{a_1a_2a_3}\epsilon^{b_1b_2b_3}\,\tr(f\,\psi_{a_1})\,\tr(\lambda^{\alpha}\,\psi_{b_1})\,\tr(\psi_{a_2}\,\psi_{b_2})\,\tr(\lambda_{\alpha}\,\psi_{a_3}\,\psi_{b_3})\nonumber \\
        &-120\,\epsilon^{a_1 a_2 a_3}\,\tr(\phi^{a}\,\psi_{a_1})\,\tr(\phi^{b}\,\psi_{a_2})\,\tr(\lambda^{\alpha}\,\psi_{(a})\,\tr(\psi_{b)}\,D_{\alpha}\psi_{a_3}\,f)\nonumber \\
        &-120\,\epsilon^{a_1 a_2 a_3}\,\tr(\phi^{a}\,\psi_{a_1})\,\tr(\phi^{b}\,\psi_{a_2})\,\tr(D^{\alpha}\psi_{a_3}\,\psi_{(a})\,\tr(\psi_{b)}\,\lambda_{\alpha}\,f)\nonumber \\
        &+20\,\epsilon^{a_1 a_2 a_3}\,\tr(\phi^{i}\,\psi_{j})\,\tr(\psi_{a_1}\,\psi_{a_2})\,\tr(\lambda^{\alpha}\,\psi_{i})\,\tr(f\,D_{\alpha}\phi^{j}\,\psi_{a_3})\nonumber \\ 
        &+20\,\epsilon^{a_1 a_2 a_3}\,\tr(\phi^{i}\,\psi_{j})\,\tr(\psi_{a_1}\,\psi_{a_2})\,\tr(D^{\alpha}\phi^{j}\,\psi_{i})\,\tr(f\,\lambda_{\alpha}\,\psi_{a_3})\nonumber \\ 
        &+20\,\epsilon^{a_1 a_2 a_3}\,\tr(\phi^{i}\,\psi_{j})\,\tr(D^{\alpha}\phi^{j}\,\psi_{a_1})\,\tr(\lambda_{\alpha}\,\psi_{a_2})\,\tr(f\,\psi_{i}\,\psi_{a_3})\nonumber \\ 
        &+20\,\epsilon^{a_1 a_2 a_3}\,\tr(\psi_{a_1}\,\psi_{a_2})\,\tr(f\,\psi_{i})\,\tr(\lambda^{\alpha}\,\psi_{j})\,\tr(\phi^{j}\,D_{\alpha}\phi^{i}\,\psi_{a_3})\nonumber \\ 
        &+20\,\epsilon^{a_1 a_2 a_3}\,\tr(\psi_{a_1}\,\psi_{a_2})\,\tr(f\,\psi_{i})\,\tr(D^{\alpha}\phi^{i}\,\psi_{j})\,\tr(\phi^{j}\,\lambda_{\alpha}\,\psi_{a_3})\nonumber \\ 
        &-20\,\epsilon^{a_1 a_2 a_3}\,\tr(f\,\psi_{i})\,\tr(\lambda^{\alpha}\,\psi_{a_3})\,\tr(D_{\alpha}\phi^{i}\,\psi_{a_2})\,\tr(\phi^{j}\,\psi_{j}\,\psi_{a_1})\nonumber \\ 
        &+20\,\epsilon^{a_1 a_2 a_3}\,\tr(\phi^{i}\,\psi_{a_1})\,\tr(f\,\psi_{a_2})\,\tr(\lambda^{\alpha}\,\psi_{j})\,\tr(D_{\alpha}\phi^{j}\,\psi_{i}\,\psi_{a_3})\nonumber \\ 
        &+20\,\epsilon^{a_1 a_2 a_3}\,\tr(\phi^{i}\,\psi_{a_1})\,\tr(f\,\psi_{a_2})\,\tr(D^{\alpha}\phi^{j}\,\psi_{j})\,\tr(\lambda_{\alpha}\,\psi_{i}\,\psi_{a_3})\nonumber \\ 
        &-28\,\epsilon^{a_1 a_2 a_3}\,\tr(\phi^{i}\,\psi_{a_1})\,\tr(f\,D^{\alpha}\phi^{j})\,\tr(\lambda_{\alpha}\,\psi_{a_2})\,\tr(\psi_{i}\,\psi_{j}\,\psi_{a_3})\nonumber \\ 
        &+28\,\epsilon^{a_1 a_2 a_3}\,\tr(\phi^{i}\,\psi_{a_1})\,\tr(f\,\lambda^{\alpha})\,\tr(D_{\alpha}\phi^{j}\,\psi_{a_2})\,\tr(\psi_{i}\,\psi_{j}\,\psi_{a_3})\nonumber \\ 
        &-4\,\epsilon^{a_1 a_2 a_3}\,\tr(\phi^{i}\,D^{\alpha}\phi^{j})\,\tr(f\,\psi_{a_1})\,\tr(\lambda_{\alpha}\,\psi_{a_2})\,\tr(\psi_{i}\,\psi_{j}\,\psi_{a_3})\nonumber \\ 
        &-4\,\epsilon^{a_1 a_2 a_3}\,\tr(\phi^{i}\,\lambda^{\alpha})\,\tr(f\,\psi_{a_1})\,\tr(D_{\alpha}\phi^{j}\,\psi_{a_2})\,\tr(\psi_{i}\,\psi_{j}\,\psi_{a_3})\nonumber \\ 
        &-16\,\epsilon^{a_1 a_2 a_3}\,\tr(\phi^{i}\,\psi_{a_1})\,\tr(f\,\psi_{a_2})\,\tr(D^{\alpha}\phi^{j}\,\lambda_{\alpha})\,\tr(\psi_{i}\,\psi_{j}\,\psi_{a_3})\nonumber \\
        &-15\,\epsilon^{a_1 a_2 a_3}\epsilon^{b_1 b_2 b_3}\,\tr(D^{\alpha}\phi^{i}\,\psi_{a_1})\,\tr(\psi_{b_1}\,\psi_{b_2})\,\tr(\psi_{i}\,\psi_{a_2})\,\tr(\lambda_{\alpha}\,\psi_{a_3}\,\psi_{b_3})\nonumber \\
        &+6\,\epsilon^{a_1 a_2 a_3}\epsilon^{b_1 b_2 b_3}\,\tr(D^{\alpha}\phi^{i}\,\psi_{a_1})\,\tr(\lambda_{\alpha}\,\psi_{b_1})\,\tr(\psi_{a_2}\,\psi_{b_2})\,\tr(\psi_{i}\,\psi_{a_3}\,\psi_{b_3})\nonumber \\
        &+9\,\epsilon^{a_1 a_2 a_3}\epsilon^{b_1 b_2 b_3}\,\tr(\phi^{i}\,\psi_{a_1})\,\tr(\psi_{b_1}\,\lambda^{\alpha})\,\tr(\psi_{b_2}\,D_{\alpha}\psi_{a_2})\,\tr(\psi_{i}\,\psi_{a_3}\,\psi_{b_3})\nonumber \\ 
        &+15\,\epsilon^{a_1 a_2 a_3}\epsilon^{b_1 b_2 b_3}\,\tr(\phi^{i}\,\psi_{a_1})\,\tr(\psi_{i}\,D^{\alpha}\psi_{b_1})\,\tr(\psi_{a_2}\,\psi_{b_2})\,\tr(\psi_{a_3}\,\psi_{b_3}\,\lambda_{\alpha})\nonumber \\ 
        &-15\,\epsilon^{a_1 a_2 a_3}\epsilon^{b_1 b_2 b_3}\,\tr(\phi^{i}\,\psi_{a_1})\,\tr(\psi_{i}\,\lambda^{\alpha})\,\tr(\psi_{a_2}\,\psi_{b_1})\,\tr(\psi_{a_3}\,D_{\alpha}\psi_{b_2}\,\psi_{b_3}) \nonumber\\
        &-18\,\epsilon^{a_1 a_2 a_3}\,\tr(\phi^{i}\,\psi_{i})\,\tr(D^{\alpha}\phi^{j}\,\psi_{j})\,\tr(\psi_{a_1}\,\psi_{a_2})\,\tr(D_{\alpha}\phi^{k}\,\psi_{k}\,\psi_{a_3})\nonumber \\ 
        &+24\,\epsilon^{a_1 a_2 a_3}\,\tr(\phi^{i}\,\psi_{a_1})\,\tr(D^{\alpha}\phi^{j}\,\psi_{i})\,\tr(\psi_{j}\,\psi_{a_2})\,\tr(D_{\alpha}\phi^{k}\,\psi_{k}\,\psi_{a_3})\nonumber \\ 
        &+12\,\epsilon^{a_1 a_2 a_3}\,\tr(\phi^{i}\,\psi_{k})\,\tr(D^{\alpha}\phi^{j}\,\psi_{i})\,\tr(\psi_{a_1}\,\psi_{a_2})\,\tr(D_{\alpha}\phi^{k}\,\psi_{j}\,\psi_{a_3})\nonumber \\ 
        &-48\,\epsilon^{a_1 a_2 a_3}\,\tr(\phi^{i}\,\psi_{j})\,\tr(D^{\alpha}\phi^{j}\,\psi_{i})\,\tr(\psi_{a_1}\,\psi_{a_2})\,\tr(D_{\alpha}\phi^{k}\,\psi_{k}\,\psi_{a_3})\nonumber \\ 
        &+96\,\epsilon^{a_1 a_2 a_3}\,\tr(\phi^{i}\,\psi_{(a_1|})\,\tr(D^{\alpha}\phi^{j}\,\psi_{|j)})\,\tr(\psi_{i}\,\psi_{a_2})\,\tr(D_{\alpha}\phi^{k}\,\psi_{k}\,\psi_{a_3}) \ .
\end{align}
}
Both $O_1^{\prime\prime}$ and $\widetilde{O}_1$ carry charges $(2,2,2,4,4)$ and $Y=9$.

As we have seen above, one combination $O_1^{\prime\prime}+\widetilde{O}_1=Q_1O_1$ of
$O_1^{\prime\prime}$ and $\widetilde{O}_1$ pairs with $O_1$ and gets lifted. 
The other combination, say $O_1^{\prime\prime}$, is shown to be a nontrivial cohomology 
till 1-loop in Section \ref{sec:hairy}. Combining this with another unlifted cohomology 
$O_1^\prime$ studied in Section \ref{sec:hairy}, we summarize their relations as follows.

The core operator $O_1$ and the hairy cohomology $O_1^\prime$ 
are both in the BMN sector, $d_{\rm BMN}=0$, have charges $j=36$ and
$(R_1,R_2,R_3,J_1,J_2)=(\frac{3}{2},\frac{3}{2},\frac{3}{2},\frac{9}{2},\frac{9}{2})$, 
and are $SU(3)_R$ singlets. The operators $Q_1O_1$ and $O_1^{\prime\prime}$ are 1 level higher than 
the BMN sector ($d_{\rm BMN}=1$), have charges $j=36$ and $(2,2,2,4,4)$,
and are also $SU(3)_R$ singlets. 
The charges of $(O_1,O_1^\prime)$ and $(Q_1O_1, O_1^{\prime\prime})$ are such that the action of 
$Q_1$ on the former could kinematically generate the latter. 
So these $4$ operators completely cancel in 
the index at $j=36$, and apparently there is no kinematic reason
for them to be protected at the quantum level. (At the classical level, different 
$d_{\rm BMN}$, or different bonus charge $Y$, forbids their pairings.)
However, at 1-loop, only two ($O_1$, $Q_1O_1$) of these four paired and became 
anomalous, leaving the other two ($O_1^\prime$, $O_1^{\prime\prime}$) unlifted.

One may ask what would be the true quantum fate of $O_1^\prime$ and $O_1^{\prime\prime}$.
Given that they are not paired by the 1-loop supercharge $Q_1$ despite the kinematic possibility,
the direct pairing is forbidden at all perturbative orders because higher order supercharges
will fail to match their $U(1)$ bonus charge $Y$.
Only non-perturbative effects may pair the two at the quantum level.
Alternatively, they may lift in two separate pairs with other cohomologies.
For example, both $O_1^\prime$ and $O_1^{\prime\prime}$ may stop being $Q$-closed at $n$-loop level,
i.e. $Q_n O_1^\prime \nsim 0$,\footnote{This condition is more complicated than
simply $Q_n O_1^\prime \neq 0$. For example for $n=2$, the condition would be that
$Q_2 \CO_0 + Q_1 \CO_1$, where $\CO_1$ is determined by $Q_1 \CO_0 + Q_0 \CO_1 = 0$,
be $Q_0$-exact. These conditions are straightforward to derive by extending \eqref{QOpert},
and we dub it the symbol $\nsim$.}
pairing with some operator at $d_{\rm BMN} = n$ and $n+1$ respectively,
because $Q_n$ raises $d_{\rm BMN}$ by $n$.
We also note that $n$ is bounded from above, because $Q_n$ reduces the number of fields $Y$
by $n-1$, meaning that for $n \geq Y$, $Q_n$ necessarily annihilates the operator.
On the other hand, neither can lift by being quantum $Q$-exact.
It is automatic for $O_1^\prime$ whose $d_{\rm BMN} = 0$.
For $O_1^{\prime\prime}$ whose $d_{\rm BMN} = 1$, the only possibility is to be $Q_1$-exact
from a cohomology with $d_{\rm BMN} = 0$.
However, it was found \cite{Choi:2023znd} that $(O_1,O_1^\prime)$ are the complete set of
the latter cohomologies in the corresponding charge sector
$(\frac{3}{2},\frac{3}{2},\frac{3}{2},\frac{9}{2},\frac{9}{2})$,
so this possibility is ruled out.
In any case, it may be interesting to see if one can leverage the bonus symmetry,
in particular the upper bound on $n$,
to argue non-renormalization phenomena to some extent.

Another possibility is that, since both $O_1^\prime$, $O_1^{\prime\prime}$ are hairy cohomologies of $O_0$
(which should be absolutely protected), somehow there is a hidden 
structure that protects such hairy cohomologies of the type 
$(O_0\textrm{ supermultiplet})\times(\textrm{gravitons})$.
This scenario is further supported by another hairy cohomology $w_2O_0$ \eqref{w2O0}
also seemingly not being lifted.
This conclusion was confirmed at 1-loop in Section \ref{sec:hairy},
and its appearance in the index under certain assumptions,
both in the $SU(2)$ \cite{Choi:2023znd} and $SU(3)$ \cite{Choi:2023vdm} theories, adds plausibility.

\section{Large charge entropy}\label{sec:entropy}

Having seen many classical black hole (fortuitous) cohomologies lifted at the 1-loop
level, one can ask by how much the classical cohomologies over-estimate the 
exact $\frac{1}{16}$-BPS states of the interacting theory. 
In particular, we want to understand whether 
the leading asymptotic entropy of the classical cohomologies at large charges is 
larger than the true $\frac{1}{16}$-BPS entropy. Although we cannot address this 
question directly, we provide a rather simple estimate which answers a related 
question from the so-called refined index of classical cohomologies.

We consider this refined index of the $U(N)$ theory. 
The $SU(N)$ index can be obtained from this by dividing that of the free $U(1)$ theory. Further setting $N=2$ will address 
the $SU(2)$ theory that we have been studying so far. The classical cohomology problem
preserves a bonus $U(1)$ symmetry, which is basically the number of letters $Y$ 
in the operator. Using this, one can study the following 5-parameter index \cite{Chang:2013fba} for the classical cohomologies,
\begin{eqnarray}
  Z(u,v,w;p,q)&=&{\rm Tr}\left[(-1)^F p^{J_1+T}q^{J_2+T}
  u^{R_1-T}v^{R_2-T}w^{R_3-T}\right]\\
  &=&\int_{U(N)} dU\exp\left[\sum_{n=1}^\infty\frac{1}{n}
  \left(1-\frac{(1-u^n)(1-v^n)(1-w^n)}{(1-p^n)(1-q^n)}\right)
  {\rm tr}(U^n){\rm tr}(U^{-n})\right]
  \nonumber
\end{eqnarray}
where $T\equiv R_1+R_2+R_3-Y$. It is an index in the sense that 
the free QFT calculation is reliable, which is the integral expression 
on the second line. The integral over $U$ can be 
reduced to that over its eigenvalues $e^{i\alpha_a}$ 
($a=1,\cdots,N$), where 
$dU\rightarrow \frac{1}{N!}d^N\alpha \prod_{a\neq b}(1-e^{i\alpha_{ab}})$ 
and $\alpha_{ab}\equiv \alpha_a-\alpha_b$. 
We shall call this the classical index, to distinguish from the 4-parameter protected index
\cite{Romelsberger:2005eg,Kinney:2005ej}.

For simplicity, we will unrefine this index as $u=v=w$, and consider 
\begin{equation}
  Z={\rm Tr}\left[(-1)^Fp^{j_1}q^{j_2}x^{R-T}\right]
\end{equation}
where $R\equiv\frac{R_1+R_2+R_3}{3}$, $j_{i}\equiv J_i+R$ and $x\equiv \frac{u^3}{pq}$.
We are further interested its Cardy limit, taking $\omega_{1,2}\rightarrow 0$ with 
$(p,q)=(e^{-\omega_1},e^{-\omega_2})$. Note that $x \to u^3$ in this limit.
As long as one is interested 
in the leading $\log Z\propto \frac{1}{\omega_1\omega_2}$ part, 
its integral expression can be approximated as 
\cite{Choi:2018hmj,Honda:2019cio,ArabiArdehali:2019tdm,Kim:2019yrz,Cabo-Bizet:2019osg}
\begin{equation}
  Z\sim \int d^N\alpha
  \exp\left[-\frac{1}{\omega_1\omega_2}
  \sum_{a,b=1}^N\left({\rm Li}_3(e^{i\alpha_{ab}})
  -3{\rm Li}_3(e^{i\alpha_{ab}}u)+3{\rm Li}_3(e^{i\alpha_{ab}}u^2)
  -{\rm Li}_3(e^{i\alpha_{ab}}u^3)\right)\right]\ .
\end{equation}
The integral can be done by a saddle point approximation for small 
$\omega_1\omega_2$. The Cardy saddle is given by all $\alpha_{ab}\approx 0$, 
i.e. all $\alpha_a$ (approximately) equal. Inserting this value into the integrand, 
the leading Cardy free energy of the classical index is given by
\begin{equation}
  \log Z\sim -\frac{N^2}{\omega_1\omega_2}
  \left({\rm Li}_3(1)-3{\rm Li}_3(u)+3{\rm Li}_3(u^2)-{\rm Li}_3(u^3)\right)
  \equiv\frac{N^2}{\omega_1\omega_2}f(\mu)\ .
\end{equation}
$\omega_i$ are the chemical potentials for $j_i$, while $u\equiv e^{-\mu}$ 
in the Cardy limit 
is that for the extra bonus charge $3(R-T)\equiv q$. 
The (protected) index capturing the exact $\frac{1}{16}$-BPS states 
is recovered by taking $x=1$, i.e. 
$u^3=1$ in the Cardy limit. 

The asymptotic micro-canonical entropy in the Cardy limit is obtained 
by the Legendre transformation, extremizing the function 
\begin{equation}\label{classical-S}
  S_{\rm cl}(j_i,q;\omega_i,\mu)=\log Z(\omega_i,\mu)+j_1\omega_1+j_2\omega_2+ q \mu
\end{equation}
in $\omega_i$, $\mu$.
This captures the indicial entropy of the classical cohomologies, refined with
$q$. As a reference, note that the indicial entropy $S$ for the protected index 
is obtained by Legendre transforming with respect to $\omega_i$ only, 
but unrefining $u^3=1$ rather than Legendre transforming it. Among the $3$ possible 
roots of the last cubic equation, it is well known that the BPS black hole entropy 
is obtained by choosing the pair of complex-conjugate roots 
$u_\pm=e^{\pm\frac{2\pi i}{3}}$, i.e. $\mu=\mp\frac{2\pi i}{3}$
\cite{Cabo-Bizet:2018ehj,Choi:2018hmj,Benini:2018ywd}. 
The resulting indicial entropy for the protected BPS states is given by
\begin{equation}\label{exact-S}
  S(j_i)=\frac{N^2f(\mp \frac{2\pi i}{3})}{\omega_1\omega_2}+j_1\omega_1+j_2\omega_2
  \ \rightarrow \ e^{\pm\frac{\pi i}{6}}
  \left[4\pi^3 N^2j_1j_2\right]^{\frac{1}{3}}~,
\end{equation}
where the last step denotes extremizing in $\omega_i$, and we used the 
fact that \cite{Choi:2018hmj}
\begin{equation}
  f\left(\mp \frac{2\pi i}{3}\right)=\frac{1}{2}\left(\pm\frac{2\pi i}{3}\right)^3
  =\mp\frac{4\pi^3i}{27}\ .
\end{equation}
We chose the solutions of the Legendre transformation which yields the positive 
${\rm Re}\,(S)$, as this is interpreted as the leading large charge entropy of 
the index.\footnote{The imaginary part of $S$ represents the sign oscillation 
$\sim \cos({\rm Im}(S)+\cdots)$ of the index \cite{Murthy:2020scj,Agarwal:2020zwm}
from the pair $u_\pm$ of choices (which actually means pair of saddles for 
the Legendre transformation).} 
The 3-parameter extremization of $S_{\rm cl}(j_i,q)$ is more complicated, 
which we shall discuss below. 

Note that both $S$ and $S_{\rm cl}$ include supergravitons and monotone
cohomologies, apart from the black hole states/fortuitous cohomologies. However, 
since $S$ and $S_{\rm cl}$ scale like $N^2$ at large $N$, 
the gravitons/monotones cannot contribute to the entropy at this order. 
Therefore, we regard the expressions above as counting the asymptotic growth 
of the protected BPS black hole states and the classical fortuitous cohomologies, 
respectively.

The classical index $e^{S_{\rm cl}(j_i,q_B)}$
refines and decomposes the protected index as 
\begin{equation}\label{relation-indices}
  e^{S(j_i)}=\sum_{q} e^{S_{\rm cl} (j_i,q)}
\end{equation}
with the bonus charge $q$. Note that all $e^{S}$ terms may be either positive/negative
integers because of nonzero ${\rm Im}(S)$. In particular, the right hand side is 
an alternating sum.
The question we would like to ask is whether there are terms on the right hand side 
which can be macroscopically larger than $e^{S(j_i)}$ on the left hand side. If this 
happens, it would mean that the sum on the right hand side exhibits substantial 
cancellations between bosonic/fermionic classical cohomologies. 
These cancellations may come from the pairs of classical cohomologies unprotected by
the classical index. For instance, the classical cohomologies lifted at 1-loop 
that we studied in Sections \ref{sec:problem} and \ref{sec:results} should all cancel this way.

Now we show that indeed there are such macroscopic cancellations. We will
show the following. After extremizing (\ref{classical-S}), we will study when the entropy
${\rm Re}\, (S_{\rm cl}(j_i,q))$ is maximized in $q$.
We will show that this maximal entropy is larger than 
${\rm Re}\, (S(j_i))$ given by (\ref{exact-S}), indeed illustrating 
a macroscopic cancellation of (\ref{relation-indices}) as discussed in the 
previous paragraph.

Let us organize the extremization/maximization problem of (\ref{classical-S})
as follows. After extremizing it in $\omega_i$ first, one obtains
\begin{equation}
  S_{\rm cl}=F(j_i,\mu)+\mu q\ \ ,\ \ 
  F(j_i,\mu)=3\left[N^2j_1j_2\right]^{\frac{1}{3}}f(\mu)^{\frac{1}{3}}\ .
\end{equation}
We then extremize it in $\mu$ to obtain $\mu(q)$ by solving $q=-\frac{\partial F}{\partial\mu}$.
Then we maximize 
\begin{equation}\label{real-entropy}
  {\rm Re}\left[S_{\rm cl}(j_i,\mu(q),q)\right]
\end{equation}
in $q$. This requires
\begin{equation}
  0=\frac{\partial}{\partial q}{\rm Re}[S_{\rm cl}(j_i,\mu(q),q)]
  ={\rm Re}\left[\mu(q)+q\frac{d\mu(q)}{dq}+\frac{\partial F(j_i,\mu(q))}{\partial\mu(q)}\frac{d\mu(q)}{dq}\right]
  ={\rm Re}\left[\mu(q)\right]
\end{equation}
where on the last step we used $q=-\frac{\partial F}{\partial \mu}$ at $\mu=\mu(q)$.
So when the entropy is maximal in $q$, its chemical potential $\mu(q)$ is purely imaginary.\footnote{A system with finite dimensional Hilbert space (e.g. Ising model) has infinite temperature $\beta=\frac{dS(E)}{dE}=0$ when its entropy is maximal, 
analogous to ${\rm Re}\, (\mu)=0$ here. Note that we are considering a micro-canonical 
ensemble with $j_1,j_2$ fixed, so the Hilbert space has finite dimension
and $q$ is bounded as $q\leq 3\min{(j_1,j_2)}$.}
With imaginary $\mu(q)$ (and of course real $q$), one obtains from (\ref{real-entropy}) the maximal entropy
\begin{equation}\label{max-entropy}
 {\rm Re}\left[F(j_i,\mu(q))+\mu(q)q\right]={\rm Re}[F(j_i,\mu(q))]
 =3(N^2j_1j_2)^{\frac{1}{3}}{\rm Re}\left[f(\mu(q))^{\frac{1}{3}}\right]\ .
\end{equation}

\begin{figure}[t!]
    \centering
    \includegraphics[width=0.6\linewidth]{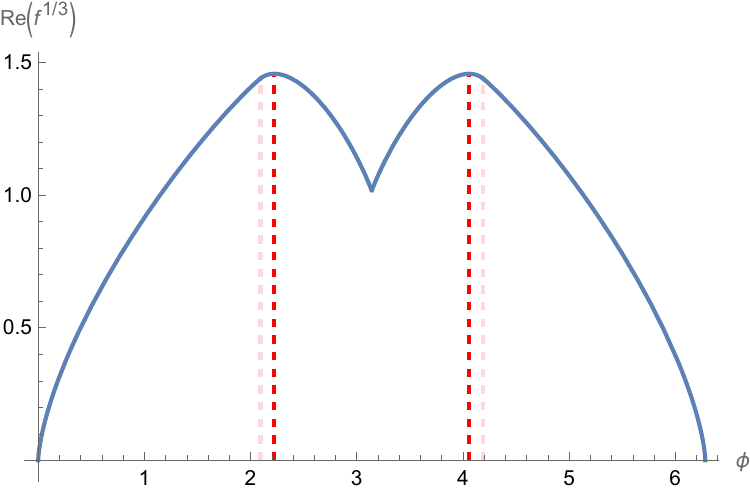}
    \caption{The plot of ${\rm Re}[(f(-i\phi))^{\frac{1}{3}}]$. It is 
    maximal at $\phi_0\approx 2.2247 = 0.7081 \pi$ and $2\pi-\phi_0$ (thick dashes), close to 
    the points $\frac{2\pi}{3}$, $\frac{4\pi}{3}$ for the 
    exact indicial entropy (light dashes).}
    \label{refined-max}
\end{figure}

In general, with a complex function $F$, $\mu(q)$ should assume a suitable complex 
value to ensure that the charge $q=-\frac{\partial F}{\partial\mu}$ is real, i.e. 
${\rm Im}(\frac{\partial F}{\partial\mu})=0$. The last equation should hold for 
arbitrary variation in the complex $\mu$ plane around $\mu(q)$. At the maximal 
entropy point, i.e. with imaginary $\mu(q)=-i\phi(q)$, we study this equation 
by varying $F$ in the real $\phi$ direction: 
\begin{equation}
  0={\rm Im}\left(i\frac{\partial F}{\partial\phi}\right)
  =\frac{\partial{\rm Im}(iF)}{\partial\phi}=\frac{\partial{\rm Re}(F)}{\partial\phi}\ .
\end{equation}
${\rm Re}(F)$ in the last expression is nothing but the maximal entropy 
(\ref{max-entropy}). So the maximal entropy can be easily found 
as the maximum of ${\rm Re}(F)\propto {\rm Re}(f^{\frac{1}{3}})$ as a function of real 
$\phi$:
\begin{equation}
  {\rm Re}\left[f(-i\phi)^{\frac{1}{3}}\right]=
  {\rm Re}\left[\left(-{\rm Li}_3(1)+3{\rm Li}_3(e^{i\phi})
  -3{\rm Li}_3(e^{2i\phi})+{\rm Li}_3(e^{3i\phi})
  \right)^{\frac{1}{3}}\right]\ .
\end{equation}
Among the three roots of $\left(\cdots\right)^{\frac{1}{3}}$, 
we choose the one with maximal real part since this yields the biggest entropy. The
plot is shown in Fig. \ref{refined-max}.

Not all part of this curve is meaningful as the entropy, because $\phi(q)=i\mu(q)$
is not real for generic $q$. It only represents certain 
entropy at two classes of points. First, the maximum of 
this curve yields the maximal entropy of the classical index.
The maximum appears at $\phi=\phi_0\approx 2.2247  = 0.7081 \pi$ and $2\pi-\phi_0$. 
Second, one can obtain the protected index (for the true $\frac{1}{16}$-BPS states)
by setting $u=e^{\pm\frac{2\pi i}{3}}$, i.e. inserting 
$\phi=\frac{2\pi}{3}$ or $\frac{4\pi}{3}$ to this function.
$\phi_0$ happens to be quite close to 
$\frac{2\pi}{3}\approx 2.0944$. The exact indicial entropy 
and the maximum of the refined indicial entropy of classical cohomologies have the following 
ratio, very close to $1$ as Fig. \ref{refined-max} shows:
\begin{equation}
  \frac{{\rm Re}[f(2.2247i)^{\frac{1}{3}}]}
  {{\rm Re}[f(\frac{2\pi i}{3})^{\frac{1}{3}}]}
  \approx 1.01213\ .
\end{equation}
The latter is larger than the former by $\sim 1.2\%$.

Based on the successful account for the BPS black hole entropy in AdS$_5$ from the index,
it is strongly believed that this index captures the leading asymptotic
entropy of the $\frac{1}{16}$-BPS states (at least in certain ranges of charges). 
It is less clear whether the refined 
index of the classical cohomologies captures their true asymptotic entropy. So 
conservatively, we interpret this ratio as a lower bound on by how much 
the classical cohomologies over-estimate the entropy of BPS states. 
Even if the over-estimate for the entropy looks quite small, this calculation 
predicts that most of the classical cohomologies at large charges 
are lifted, 
i.e.
\begin{equation}
  e^{S_{\rm cl}(j_i)}\sim e^{1.012 S(j_i)}\gg e^{S(j_i)}\ .
\end{equation}
That is, the fraction of surviving classical cohomologies is 
no bigger than $e^{-0.012 S(j_i)}\ll 1$.

One may still wonder why the over-estimation of the entropy by classical 
cohomologies (as seen by the refined index) is so small. 
This is in contrast to the ratio of the free BPS operators' entropy vs. the indicial entropy, 
say in the Cardy limit, which is substantially larger than $1$. Perhaps the 
difference might be related to how many accidental symmetries appear in the free 
QFT and in the interacting classical cohomologies (1-loop Hamiltonian), relative to 
the generic interacting theories. The free theory has infinite dimensional higher 
spin symmetries, while the only accidental symmetry seen in the 1-loop Hamiltonian
is the bonus $U(1)_Y$ symmetry. We are suspecting this possibility because the 
over-estimation probed in this section used the $U(1)_Y$ symmetry, 
but it would be nice to understand this more precisely.

\vskip 0.5cm

\hspace*{-0.8cm} {\bf\large Acknowledgements}
\vskip 0.2cm

\hspace*{-0.75cm} This work is supported in part by the NRF grant RS-2026-25477643 
(Seok Kim, Seongmin Kim, JP),
FWO projects G094523N and G003523N, KU Leuven project C16/25/010 and the FWO Junior Postdoctoral
Fellowship 1274626N (SL).
We thank Eunwoo Lee and Jehyun Lee for helpful discussions.

\appendix
\section{Linear algebra over a finite field $\mathbb{Z}_p$}\label{app:modp}

In this appendix, we explain the technique of converting all numbers into a finite field $\mathbb{Z}_p$,
used to boost arithmetics for many linear algebra problems.

As we have discussed towards the end of Section \ref{sec:problem},
the problem of determining $Q_0$-exactness of an operator becomes that of
solving a linear system with a very large matrix.
During the computations, for example under repeated Gaussian eliminations, the entries grow rapidly,
and either the exact arithmetic becomes unbearable or the floating-point arithmetic
introduces round-off ambiguities. To avoid both problems, one can carry out the entire computation over $\mathbb{Z}_p$, the finite field with $p$ elements~\cite{BoroshFraenkel1966}, by
\begin{align}
x \bmod p \;\equiv\; x - p\left\lfloor \tfrac{x}{p} \right\rfloor \quad (x\in\mathbb{Z})\,,
\qquad
\mathbb{Z}_p \equiv \{0,1,\dots,p-1\}\,.
\end{align}
Choosing $p < 2^{31}$ guarantees that the product of any two reduced elements satisfies $a\cdot b < 2^{62} < 2^{63}$, so every intermediate quantity fits in the signed 64-bit integer type without overflow. A rational number $a/b$ with $\gcd(b,p)=1$ is embedded into $\mathbb{Z}_p$ by
\begin{align}
\frac{a}{b} \;\longmapsto\; a\,b^{-1} \bmod p\,,
\end{align}
where $b^{-1}$ refers to the modular inverse that follows from Fermat's little theorem,
\begin{align}
x^{p-1} \equiv 1 \pmod p
\quad\Longrightarrow\quad
x^{-1} \equiv x^{p-2} \pmod p
\qquad (\gcd(x,p)=1)\,.
\end{align}
This map $\mathbb{Q}\to\mathbb{Z}_p$ is many-to-one, but it becomes invertible on rationals of sufficiently small denominator and numerator: for each $x\in\mathbb{Z}_p$ there is at most one pair $(a,b)\in\mathbb{Z}\times\mathbb{Z}_{>0}$ with
\begin{align}
a \equiv b\,x \pmod p\,,
\qquad \gcd(a,b)=1\,,
\qquad |a|,\,b \le \sqrt{p/2}\,.
\end{align}
This pair is recovered by applying the extended Euclidean algorithm to $(p,x)$, a procedure known as rational reconstruction~\cite{WangGuyDavenport1982}.
Initializing
\begin{align}
(r_0,t_0) = (p,0)\,, \qquad (r_1,t_1) = (x,1)\,,
\end{align}
and iterating Euclidean division for $k\ge 1$,
\begin{align}
q_k = \left\lfloor \frac{r_{k-1}}{r_k} \right\rfloor \,,
\qquad
r_{k+1} = r_{k-1} - q_k\,r_k\,,
\qquad
t_{k+1} = t_{k-1} - q_k\,t_k\,,
\end{align}
every step preserves the congruence
\begin{align}
r_k \equiv t_k\,x \pmod p .
\end{align}
The remainders $|r_k|$ strictly decrease, so there is a first index $k$ with $|r_k| < \sqrt{p/2}$; provided $|t_k| \le \sqrt{p/2}$ as well, the reduced, sign-normalized pair $(a,b) = (r_k,t_k)/\gcd(r_k,t_k)$ with $b>0$ is exactly the pair above. If this bound is too small to pin down the original rational, the Chinese remainder theorem~\cite{KnuthTAOCP2} reconstructs it over a wider range by combining residues from several primes.

In principle, two types of error could arise from working over $\mathbb{Z}_p$ rather than $\mathbb{Q}$.
\begin{eqnarray}
    &&\text{False positive: } A\vec{x} = \vec{b} \text{ over } \mathbb{Z}_p
    \text{ has a solution but } A\vec{x} = \vec{b} \text{ over } \mathbb{Q}\text{ does not.} \nn\\
    &&\text{False negative: } A\vec{x} = \vec{b} \text{ over } \mathbb{Z}_p
    \text{ has no solution but } A\vec{x} = \vec{b} \text{ over } \mathbb{Q} \text{ does.} \nn
\end{eqnarray}
They can be controlled by separate mechanisms.
\begin{itemize}
    \item \emph{False positive.} When the Gaussian elimination over $\mathbb{Z}_p$  finds a solution, we obtain $\vec{x}\bmod p$ explicitly. The rational reconstruction then yields a candidate $\vec{x}\in\mathbb{Q}^K$, valid whenever each component has numerator and denominator below $\sqrt{p/2}\approx 3.3\times 10^4$. The candidate is confirmed by checking $A\vec{x}=\vec{b}$ over $\mathbb{Q}$.
    \item \emph{False negative.} It is possible that some independent set of $Q_0$-exact basis over $\mathbb{Q}$ becomes dependent over $\mathbb{Z}_p$. This happens when the $\mathrm{rank}\ A$ over $\mathbb{Z}_p$ is strictly less than that over $\mathbb{Q}$, which generally satisfy~\cite{BoroshFraenkel1966}
    \begin{eqnarray}
        \mathrm{rank}_{\mathbb{Z}_p}(A) \leq \mathrm{rank}_{\mathbb{Q}}(A).
    \end{eqnarray}
    The strict inequality holds precisely when $p$ divides every $r \times r$ submatrix of $A$ (with $r = \mathrm{rank}_{\mathbb{Q}}(A)$), equivalently when $p$ divides the largest invariant factor $d_r$ of $A$. The heuristic probability of this event is suppressed by (negative) powers of $p$. Repeating the computation with distinct primes $(2^{31}-1,\,2^{31}-19,\,2^{31}-61,\,2^{31}-69)$ further suppresses the false-negative probability.

\end{itemize}

\end{document}